# The use of a novel gradient heat flux sensor for characterization of reflux condensation

Filip Janasz[a,*,1], Horst-Michel Prasser[a], Detlef Suckow[b], Andrey Mityakov[c]
[a]ETH Zürich, Leonhardstrasse 21, 8092 Zürich, Switzerland
[b]Paul Scherrer Institut, Villigen 5232, Switzerland
[c]Peter the Great St. Petersburg Polytechnic University, Russia
(Email addresses: filip.janasz@uni.lu, prasser@lke.mavt.ethz.ch, detlef.suckow@psi.ch, mitiakov@spbstu.ru)

**ABSTRACT**
In this paper we present the developments in heat flux measurements using gradient heat flux sensors (GHFS). The GHFS is a sensor made of artificially created material with anisotropic thermo-electrical properties. Its properties including small size, robustness and low response time make it a valuable addition for characterizing heat flux in space-limited geometries as well as harsh environmental conditions. This makes it ideal for investigation of reflux condensation in the steam generator tubes of a pressurized water reactor. The reflux condenser mode of operation during accident as well as maintenance conditions can provide a significant passive cooling, removing the residual decay heat from the reactor core, thus preventing or at least delaying potential core uncovery. With this novel implementation of GHFS deeper characterization of heat flux during reflux condensation was achieved.

**NOMENCLATURE**

**Abbreviations**
| | |
|---|---|
| $NC$ | non-condensable |
| $PWR$ | pressurized water reactor |
| $SG$ | steam generator |
| $TC$ | thermocouple |

**Latin letters**
| | |
|---|---|
| $A$ | area |
| $E$ | GHFS voltage response |
| $\dot{m}$ | mass flow |
| $\dot{Q}$ | heat flux |
| $S$ | Seebeck coefficient |
| $s_0$ | GHFS sensitivity |
| $T$ | temperature |

**Subscripts**
| | |
|---|---|
| $wall$ | condensing tube wall |
| $in$ | inner wall surface |
| $out$ | outer wall surface |
| $cond$ | condensation |
| $ev$ | evaporation |
| $el$ | electric |
| $err$ | error |

---

[1] Present address: Faculty of Science, Technology and Medicine, Department of Engineering, Campus Kirchberg, University of Luxembourg, 6, rue Richard Coudenhove-Kalergi, L-1359 Kirchberg, Luxembourg

## 1 INTRODUCTION

Reflux condensation can occur in steam generator (SG) tubes of a pressurized water reactor (PWR) during accidents or mid-loop operation. In this scenario, SG tubes are filled with steam or steam / non-condensable gas mixture instead of water. The reactor core is still at least partially submerged, and the hot steam produced by the decay heat is carried upwards to the SG tube bundle. A portion of it condenses and returns to the reactor core, thus the counter-current gas and liquid flow regime is present in the reactor's hot leg and SG. As a result, a possibly significant passive residual decay heat removal from the reactor's primary coolant system is established [1–3]. Reflux condensation can occur at high pressures 4-10 MPa, due to a loss of coolant accident (LOCA) during normal plant operation [4–8] or at containment pressures due to loss of residual heat removal (RHR) system during a mid-loop inventory operation [9–11].

During both scenarios, presence of non-condensable (NC) gases is likely. For LOCA-related cases, possible sources include the pressurizer ($N_2$), oxidizing fuel rods zirconia cladding ($H_2$) and gases normally dissolved in the coolant water [12]. Alternatively, during mid-loop operation, the primary coolant system is intentionally partially drained, and air is present in the SG tubes.

The Precise Reflux Condensation Investigation Setup (PRECISE) facility was developed at Paul Scherrer Institut to study the interaction of steam and non-condensable gases in the geometry of PWR SG tubes. It is a vertical concentric tube condenser, with the ability to control the gas inventory and operating conditions such as pressure, temperature and coolant properties. To expand the measurement scope, besides typical measurement methods – thermocouples and pressure transducers – the facility is equipped with novel gradient heat flux sensors, developed by and implemented in cooperation with Mityakov et. al [13,14]. Their small thickness and thermal properties allowed to introduce very little disturbance to the heat flux through the condensing tube wall. Since the key feature of interest of reflux condensation is its ability to remove the residual decay heat, accurate characterization of the observed heat flux is of significant value. GHFS allowed to observe it with previously unavailable spatial and temporal resolution.

## 2 PRECISE FACILITY AND EXPERIMENTS DESIGN

The PRECISE facility is made of two main subsystems – the primary loop, in which steam is produced, transported to the test tube and condensed, and the secondary loop, which supplies and controls the steady flow of coolant around it. The primary loop includes the steam generation tank and the condensation tube, while the secondary loop includes a cooling jacket, water storage tank, pump and heat exchanger. The cross-sectional overview of the main body of the facility and its implementation is shown in Figure 1. The test rig is additionally equipped with water and NC gases feeding tanks, which allow for a precise control of the gas inventory in the test tube during experiment setup. Nitrogen and helium (in place of flammable hydrogen) are used, as well as mixtures of the two at different ratios.

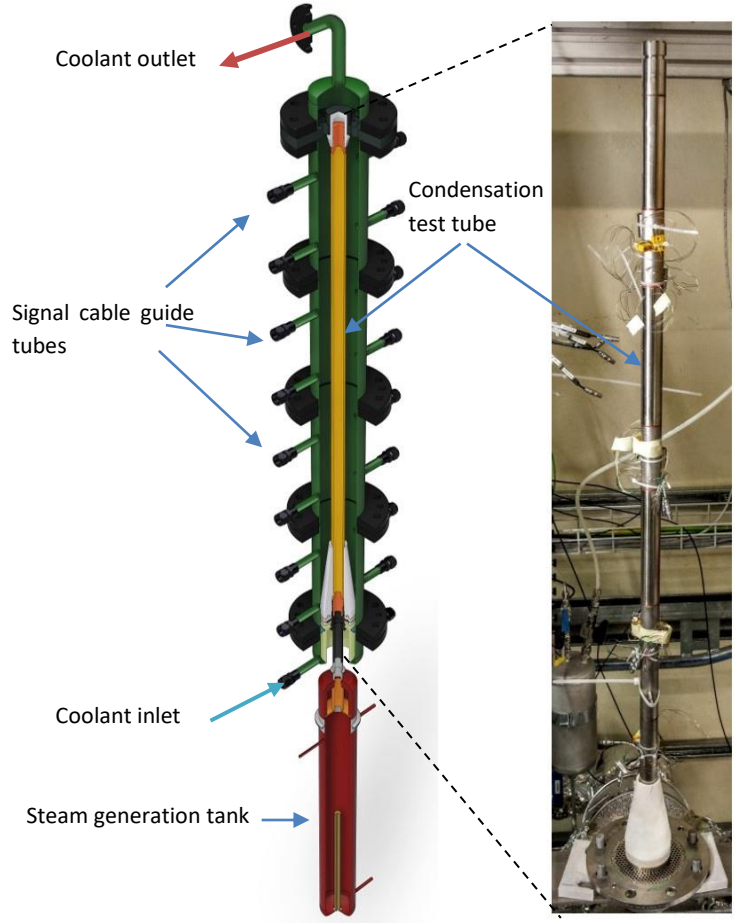

**Figure 1 PRECISE facility cross-section (left) and the condensation test tube.**

The condenser tube has 20 mm inner diameter representative of PWR steam generator tubes geometry and the total height of 1300 mm. 5 mm thick wall allows for placement of GHFS and the development of temperature gradient for differential temperature measurement. The added mass in the thick tube wall increased the thermal inertia of the overall system. However, since experiments were conducted during established steady-state or slowly changing conditions, and heat flux sensors were placed very close (0.5 mm) to the wall inner surface, this did not significantly affect the measurements. The test tube is closed at the top and forms a single volume with the steam generation tank located below. Steam is produced with an electrical heating rod (up to 2 kW), located in the generation tank. Pressure measurement was used as a feedback parameter to control the heater. In this arrangement, at constant pressure and temperature, the flow of produced water vapor $\dot{m}_{ev}$ can be calculated based on the electrical power consumed by the heater $P_{ev}$ and can be considered equal to the condensation mass flow $\dot{m}_{cond}$, eq. (1):

$$P_{ev}/(H_v - H_l) = \dot{m}_{ev} = \dot{m}_{cond} \qquad (1)$$

With $H_v - H_l$ being the evaporation enthalpy of water. This reduces the uncertainties in condensation mass flow estimation, by ensuring the inherent consistency in the supplied heat and the thermal energy removed by condensation. Summary of the PRECISE facility operating conditions is presented in the Table 1. The 10 bar maximum operating pressure was set due to practical limitations of the experimental facility.

| Parameter | Value range | Unit | Instrument | Meas. error |
|---|---|---|---|---|
| *PRIMARY LOOP* | | | | |
| *Electrical heating power* | 0 – 2 | kW | Power controller | 1.41% |
| *Pressure* | 1 – 10 | bar | Pressure transducer | 1.41% |
| *Temperature* | 20 – 180 | °C | PT100 / K-type thermocouple (⌀ 0.5mm) | ±0.0316 K / ±1.006 K |
| *Steam vapor density* | 0.59 – 5.147 | kg/m$^3$ | Derived | 0.01 - 0.23% |
| *Steam mass flow* | 0 – 1 | g/s | Derived | 1.42% |
| *Steam velocity* | 0 – 5 | m/s | Derived | 1.42 - 1.44% |
| *Steam Reynolds number* | 0 – 25000 | - | Derived | |
| *Condensate mass flow* | 0 – 1 | g/s | Derived | 1.42% |
| *NC gas mole fraction* | 0 – 1 | - | Derived | 3.19 - 7.04 % |
| *NC gas species* | N$_2$ & He | - | N/A | - |
| *NC gas – N$_2$ to He ratio* | 0 – 1 | - | Derived | 1.97% - 2.16% |
| *Safety pressure release* | 12 | bar | N/A | - |
| *SECONDARY LOOP* | | | | |
| *Liquid* | Demin. water | - | N/A | |
| *Pressure* | 1 – 10 | bar | Pressure transducer | 1.41% |
| *Temperature* | 20 – 180 | °C | PT100 resistive thermometer | ±0.0316 K |
| *Water volumetric flow rate* | 0 – 5 | m$^3$/h | Vortex flowmeter | 0.016 to 0.058 m$^3$/h |
| *Water velocity* | 0 – 0.175 | m/s | Derived | 1.07% |
| *Water Reynolds number* | 0 – 70000 | - | Derived | 0.1% |
| *Safety valve release pressure* | 12 | bar | N/A | - |
| *ENVIRONMENT* | | | | |
| *Lab temperature* | 15-25 | °C | K-type thermocouple | |

**Table 1 PRECISE facility operating conditions summary**

## 3   GRADIENT HEAT FLUX SENSORS

### 3.1   Operating principle

GHFS installed in the PRECISE facility are made of an artificially produced material with anisotropic electrical and thermal properties. It is comprised of alternating layers of two materials with different Seebeck coefficients. Material layers are fused together with a diffusion welding process and the resulting structure is cut, so that the layers are tilted at a certain angle, Figure 2. The result is a multi-layered, solid structure. Due to the material anisotropy, heat flux applied perpendicularly to the sensor induces proportional transverse thermoelectric voltage response parallel to the surface. This effect is called the transverse Seebeck effect. The resulting voltage can be measured across the sensor area, in the direction perpendicular to the tilted layer structure, Figure 2.

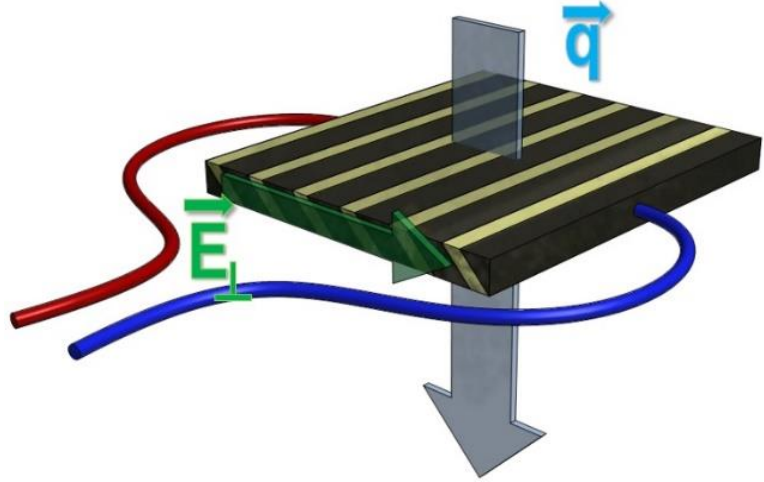

**Figure 2 Heat flux sensor multi-layer structure and operating principle**

The overall thermoelectric response $E$ of an anisotropic element is described by eq. (2):

$$\hat{E} = \hat{S}\, \nabla T \qquad (2)$$

where $\nabla T$ is the temperature gradient and $\hat{S}$ is the Seebeck coefficients tensor. Detailed explanation of the GHFS operating principle can be found in Mityakov et al. [13]. The thickness and angle of the layers can be optimized to achieve the highest possible thermoelectric response [15,16].

### 3.2 GHFS types

Any pair of materials that can be bonded and processed to produce the previously described structure could in principle act as a GHFS. Alternatively, naturally occurring anisotropic crystals can and were successfully used in creating heat flux sensing elements [13]. The choice of a certain material pair for GHFS is dictated by the required operating conditions and acceptable signal amplitude. If only the sensor response could be considered, the choice of materials pair should maximize the anisotropy, which can be defined as:

$$\Delta S = S_{\hat{x}\hat{x}} - S_{\hat{y}\hat{y}} \qquad (3)$$

where $S_{\hat{x}\hat{x}}$ and $S_{\hat{y}\hat{y}}$ refer to Seebeck coefficient tensor components parallel and perpendicular to the sensor area and are a function of constituent materials properties [15]. Table 2 shows summary of the observed values in some of previously tested materials and material pairs.

| GHFS material | $\Delta S$ [µV/K], 300K | |
|---|---|---|
| Bismuth | 50 | ([16]) |
| Copper – constantan | 35 | ([17]) |
| Constantan – chromel | 5 | ([16]) |
| Aluminum - silicon | 1500 | ([15]) |
| Stainless steel – nickel | 40 | ([18]) |
| Brass - steel | 12 | ([19]) |

**Table 2 Overview of thermoelectric anisotropy of various materials**

Based only on the GHFS response to heat flux, the Al-Si sensors seem to offer the best performance. However, in PRECISE facility stainless steel (18% Cr, 9% Ni, 2% Mn, 0.8% Ti) and nickel GHFSs were used instead. As the condensation test tube is made of stainless steel, this material composition closely resembles its properties and minimizes the disturbance to the condensation heat flux introduced in the system. Other influential factors included their resistance to corrosion and high operating temperatures, which fit the possible exposure to superheated steam and condensate water. All-metal sensors are also easy to shape and bend, which allowed matching the curvature of the tube inner wall. Their properties can be summarized as follows:

- Working temperature up to 1300 K
- Sensitivity up to 0.4 mV/W
- Response time estimated in the range of $10^{-9}$ s [14]
- 0.2 - 0.3 mm sensor thickness
- 95-100 mm² area
- <0.1 mm constituent layer thickness

The fast response time permitted GHFS response sampling rates up to 100 kHz, which, in turn, allowed to use signal super-sampling to filter the noise caused by low signal amplitude.

Ultimately four sensors were mounted in the test tube, distributed along its height in locations A – D, Figure 3. At these locations, pairs of thermocouples (type K, D=0.5 mm) are embedded in the wall for temperature gradient measurement. The listed elevation is from the entrance to the condensation tube. Besides GHFS and thermocouple pairs, additional 12 thermocouples (type K, D=0.5 mm) are located in the tube's center at the marked locations, to observe the vertical temperature distribution.

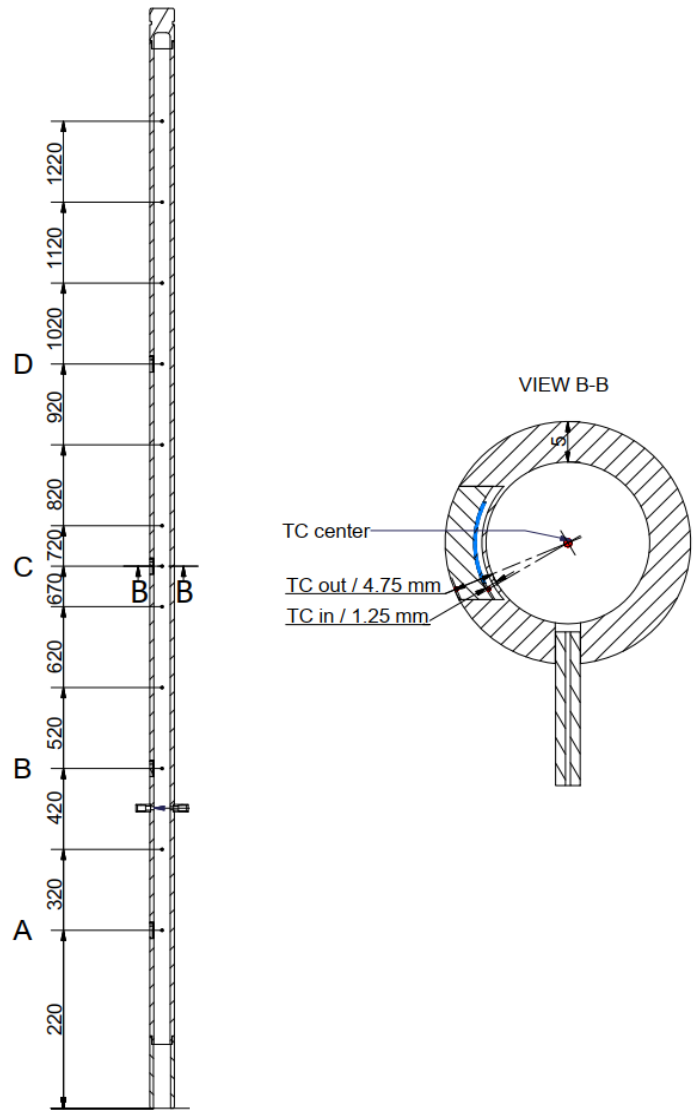

Figure 3 GHFS and thermocouple distribution in PRECISE facility.

## 3.3 GHFS wall embedding

GHFS were mounted in the condensation tube wall, behind a thin layer of steel, protecting it from a direct exposure to steam. This was necessary as proven by previous tests in which sensor was placed on the inner surface of the test tube and embedded into epoxy bed in the machined groove. Over the course of approximately two months, harsh conditions disintegrated the epoxy and necessitated modification of the mounting technique.

The steel layer thickness was 0.5 mm to limit the heat flux signal delay at the sensor location. To achieve this, a groove was machined from the outside the tube using electric-discharge die-sinking, Figure 4.

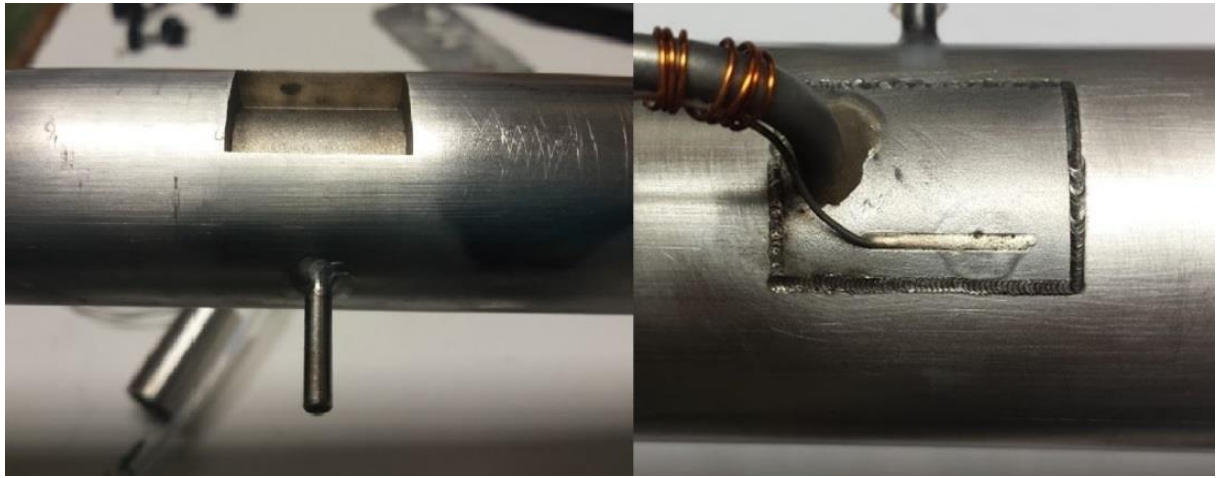

Figure 4 GHFS installation in the deep pocket. Figure reproduced from [21].

GHFS's were bent to match the curvature of the manufactured pocket and placed inside. Cover was then fitted on top and laser-welded in place to create a pressure seal. The pocket was filled with electrically insulating / thermally conductive epoxy[2] to fill the remaining empty volume. The thermal conductivity of epoxy is listed at 2.88 W/m·K, compared to 16.3 W/m·K for 1.4404 stainless steel tube wall. Filling of the void ensured good thermal contact between the wall and GHFS and reduced the local heat flux disturbance. To assess the disturbance to the temperature field in the tube wall due to the sensor placement, thermal FEM calculations using ANSYS software were conducted. No significant impact to the temperature distribution was observed – an example of calculation results with boundary conditions on both sides of the wall set to convective and free flow temperature difference set to 20 °C is shown in the Figure 5.

---

[2] Cotronics Duralco 128

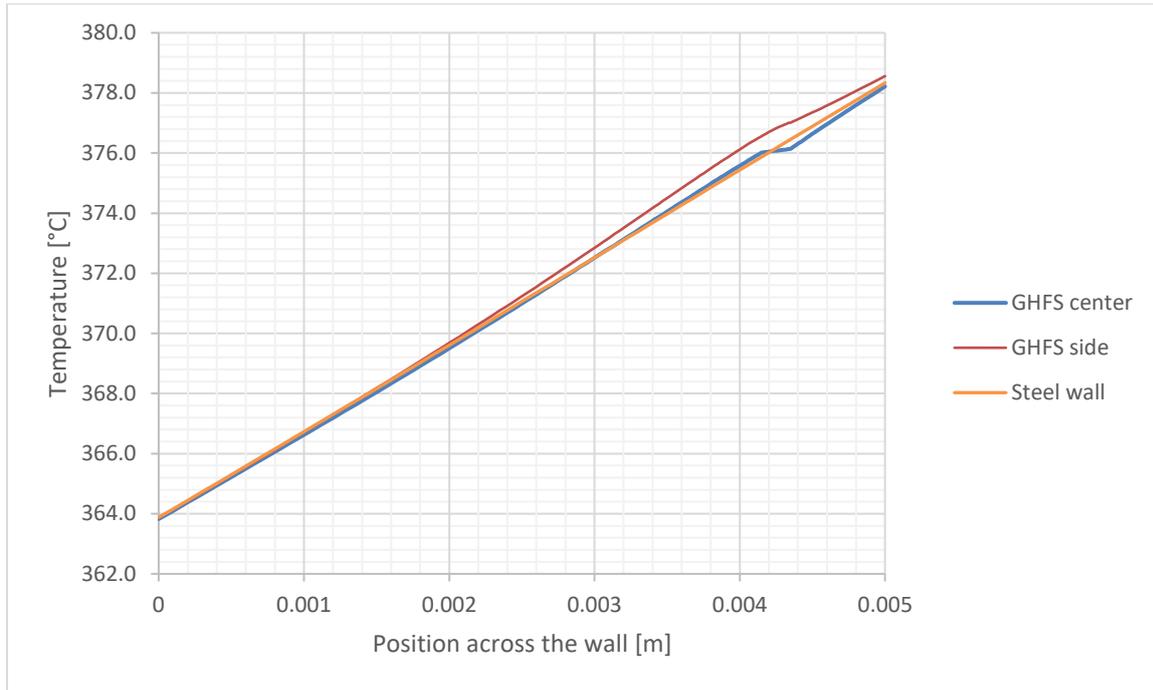

**Figure 5 Temperature distribution across the tube's wall estimated with ANSYS Fluent at three different locations.**

Sensor signal cables were evacuated through a penetration in the top cover and further through a pressure-sealed tubing. In addition to GHFS, two thermocouples (type K, D=0.5 mm) were embedded in each assembly – one in the pocket and one on the cover's surface. In Figure 4, the top thermocouple is visible under a layer of high-temperature yellowish solder. Extensive testing of this embedding method verified its temperature and pressure resistance, suitable to use in PRECISE facility.

Protecting sensors with a 0.5 mm steel layer increased the assembly thermal inertia and thus formed a low-pass filter for higher frequency heat flux phenomena. To quantify the degradation from the claimed $10^{-9}$ s response time, a small test was designed in which the embedded sensor was periodically exposed to a heat source with an adjustable exposure frequency. The setup consisted of a quartz lamp with a rotary disc shutter with a controlled shutter angle, separating it from the covered sensor. Gathered data revealed that frequencies of heat flux variation are still observable with the GHFS up to 10 Hz.

## 4 GHFS SIGNAL ACQUISITION

One limitation of a stainless steel-nickel composite GHFS is the relatively low amplitude of the electric response. The condensation heat flux in PRECISE facility was estimated to reach values up to 10000 W/m$^2$. Therefore, considering the theoretical sensor sensitivity of 0.4 mV/W the expected signal was in the range of 0-400 µV. Data logging hardware used for signal acquisition was a National Instrument voltage input module NI-9215. This module is a four-channel device with a A/D converter, capable of simultaneous and continuous signal sampling up to 100 kHz with 16-bit accuracy of each channel. The summary of its properties is listed in Table 3.

| Name | Manufacturer | Accuracy reading | Gain error | Offset error | Resolution | Resolution A/D | Gain drift | Offset drift |
|---|---|---|---|---|---|---|---|---|
| NI 9215 | National Instruments | N/A | ± 0.02% | 0.014% 1.456 mV | 0.153 mV | 16 bit | 10 ppm/K | 60 µV/K |

**Table 3 NI-9215 DAS card properties summary**

Considering the resolution, and the measurement range of ± 10 V, it was decided to include amplifier into the measurement chain. Overall, two amplification stages were used, each with gain of 100, making the total amplification factor of 10000 and increasing the expected signal range to 0 to 4 V, post amplification. The minimize the effect of noise amplification, the first amplification stage is implemented with an instrumentation amplifier characterized by high common-mode noise rejection. Additionally, this stage was powered from a LiPOs battery, to further minimize the power source noise. The overview of the full signal acquisition is shown on Figure 6.

The NI-9215 card was installed in a cRIO-9022 embedded real-time controller frame which buffered and relayed the sampled signal to the laboratory PC over Ethernet connection. A LabVIEW program was used to acquire the data.

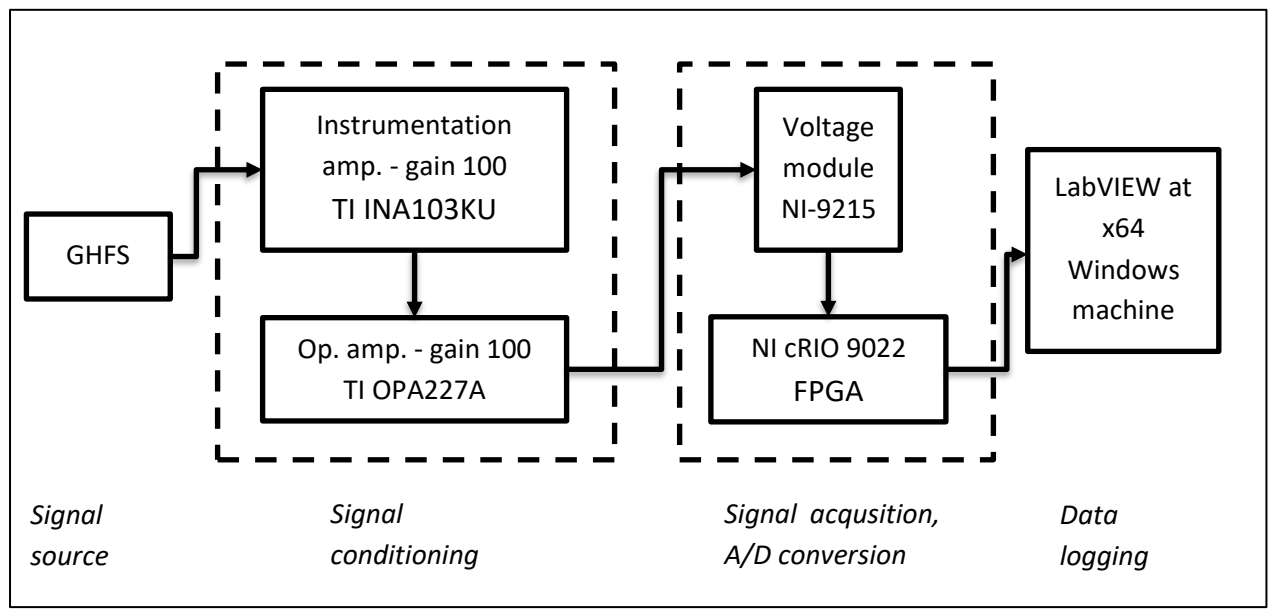

Figure 6 Signal acquisition system for GHFS sensors. Figure reproduced from [21].

## 5 MEASUREMENT ACCURACY

To convert the voltage reading of the sensor, the following equation is applied:

$$\dot{Q} = \frac{E}{A \cdot s_0} \quad (4)$$

Where $\dot{Q}$ is the heat flux applied to the sensor, $E$ is the measured output of the sensor, $A$ is the sensor area and $s_0$ is GHFS sensitivity. Each of the three parameters on the right side of equation (4) can contribute to the total measurement error due to their own uncertainties. For the purpose of overall measurement uncertainty estimation, it is assumed that they can be treated as independent statistical errors, and the total uncertainty $Q_{err}$ can be estimated according to the equation (5):

$$\dot{Q}_{err} = \sqrt{E_{err}^2 + A_{err}^2 + s_{0\_err}^2} \quad (5)$$

where $E_{err}$ is the voltage signal acquisition error, attributable to the data acquisition, $A_{err}$ is the sensor area measurement error and $s_{0\_err}$ is the error of sensitivity factor calculated via calibration procedure. Based on the NI-9215 properties, the signal range, and accuracy of the two-stage amplifier, $E_{err} \approx 2.24\%$. GHFS area was measured using a Leica DFC320 digital camera attached to a Leica MZ16 stereomicroscope. The resolution of the optical system was 9 μm and the sensor area was calculated by software. The $A_{err} = 0.45\%$ was estimated based on the error attributable to uncertainties of sensor edge tracking. Finally, sensors were calibrated using electrical heater in vacuum, supplying well-controlled heat flux $\dot{Q}_{rad}$ via radiation and measuring each GHFS response. Based on the formula (4), expression for sensitivity can be written as:

$$s_0 = \frac{E}{A \cdot \dot{Q}_{rad}} \qquad (6)$$

and the accompanying error, similarly to equation (5) is expressed as:

$$s_{0\_err} = \sqrt{E_{err\_2}^2 + A_{err}^2 + \dot{Q}_{rad\_err}^2} \qquad (7)$$

where $E_{err\_2}$ is the error attributable to the digital multimeter used during the calibration process (Keithley Model 2001, resolution 0.1 μV) and can be estimated as 0.1 % according to the data sheet. $A_{err}$ value is already known and $\dot{Q}_{rad\_err}$ was estimated based on the characteristics of the power controller (JUMO TYA 201) used to deliver the electrical power and manufacturing tolerances of the condensation test tube. Final value of $s_{0\_err}$ was estimated at 1.57%. Since this parameter is temperature dependent, error of GHFS temperature measurement by an accompanying calibrated thermocouple increases the final measurement error to 1.72%.

Finally, heat flux measurement relative error can be estimated based on equation (5) as:

$$\dot{Q}_{err} = \sqrt{1.72\%^2 + 0.45\%^2 + 2.24\%^2} \approx 2.86\% \qquad (8)$$

## 6 HEAT FLUX MEASUREMENTS

### 6.1 Steady-state tests

In this type of tests, a predefined amount of NC gas was introduced to the condensation tube, a set-point pressure was defined, and the heating of the test tube and the steam generation tank was started. After some time (40-60 minutes), depending on the chosen parameters as well as the coolant temperature / velocity, a steady-state was reached, at which the evaporation and condensation heat fluxes would equalize and pressure in the test tube remained constant. Data was collected during this period and the conditions stability was later confirmed in data post-processing. The criterion used for steady-state identification were the test-tube pressure and steam temperature stability around the average value within ± 0.075%. Data presented in this section was collected at 4 bar pressure, 20 °C wall ΔT at inlet and 2 m³/h coolant water flow. The inventory of NC gases was varied in the amount of added gas and its type.

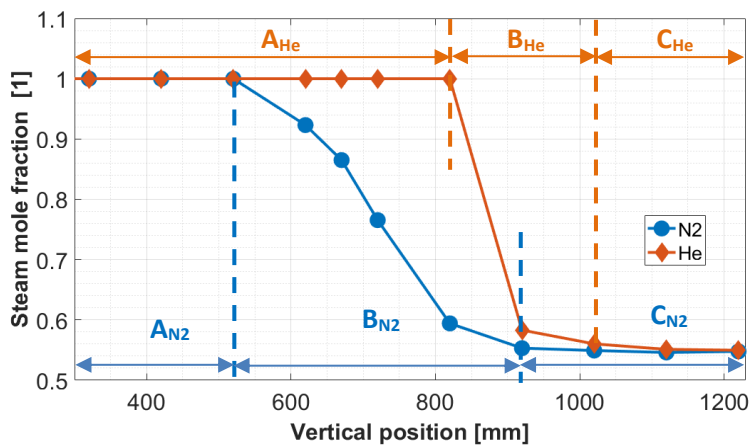

**Figure 7 Typical steam mole fraction profile along test tube centerline with three distinct regions A, B & C for experiments with $N_2$ and He. Figure reproduced from [21].**

It was observed that for both used NC gas species – nitrogen and helium, significant gas stratification was present in the condensation tube (more on this behavior can be found in Janasz et al. [20]). Therefore, three significantly different zones could be identified inside the test tube based on the water vapor mole fraction (calculated based on recorded temperature and assumption of thermodynamic equilibrium at stable pressures). Figure 7 presents the steam mole fraction distribution along the condensation tube for tests with $N_2$ (in blue) and He (in red). A is the pure steam region, starting at the inlet of the tube, characterized by high condensation rates. B is the intermediate mixing zone, where interaction between NC gas and steam flow occurs. Finally, C zone reaching all the way to the tube's top is a NC gas plug with high NC molar fraction. No condensation is observed in zone C. In each of the zones, significantly different behavior in condensation heat flux was observed with GHFS. First of all, the amplitude of the observed signal dropped continuously in the mixing zone, from the maximum values observed in steam-rich zone to almost zero in the NC gas plug on the top. Second, the oscillations in the signal were most pronounced in the zone B when compared to two other zones. These tendencies were observed for each gas species, as well as their mixtures, used in the experiments. However, as shown in the Figure 7, light gas formed much sharper interfaces than heavy gas. The molar gas distribution was followed closely by the observed heat flux.

In each of the identified zones, the heat flux signal was analyzed to observe the time-resolved, dynamic properties, Figure 8. Here, when comparing peak-to-peak values of recorded heat flux signals, the oscillations were greatest in the mixing region, followed by significantly smaller values in the condensation zone and NC plug. Moreover, in the mixing region, the oscillation amplitude was also dependent on the type of the gas species present, with $N_2$ exhibiting values twice as large as He.

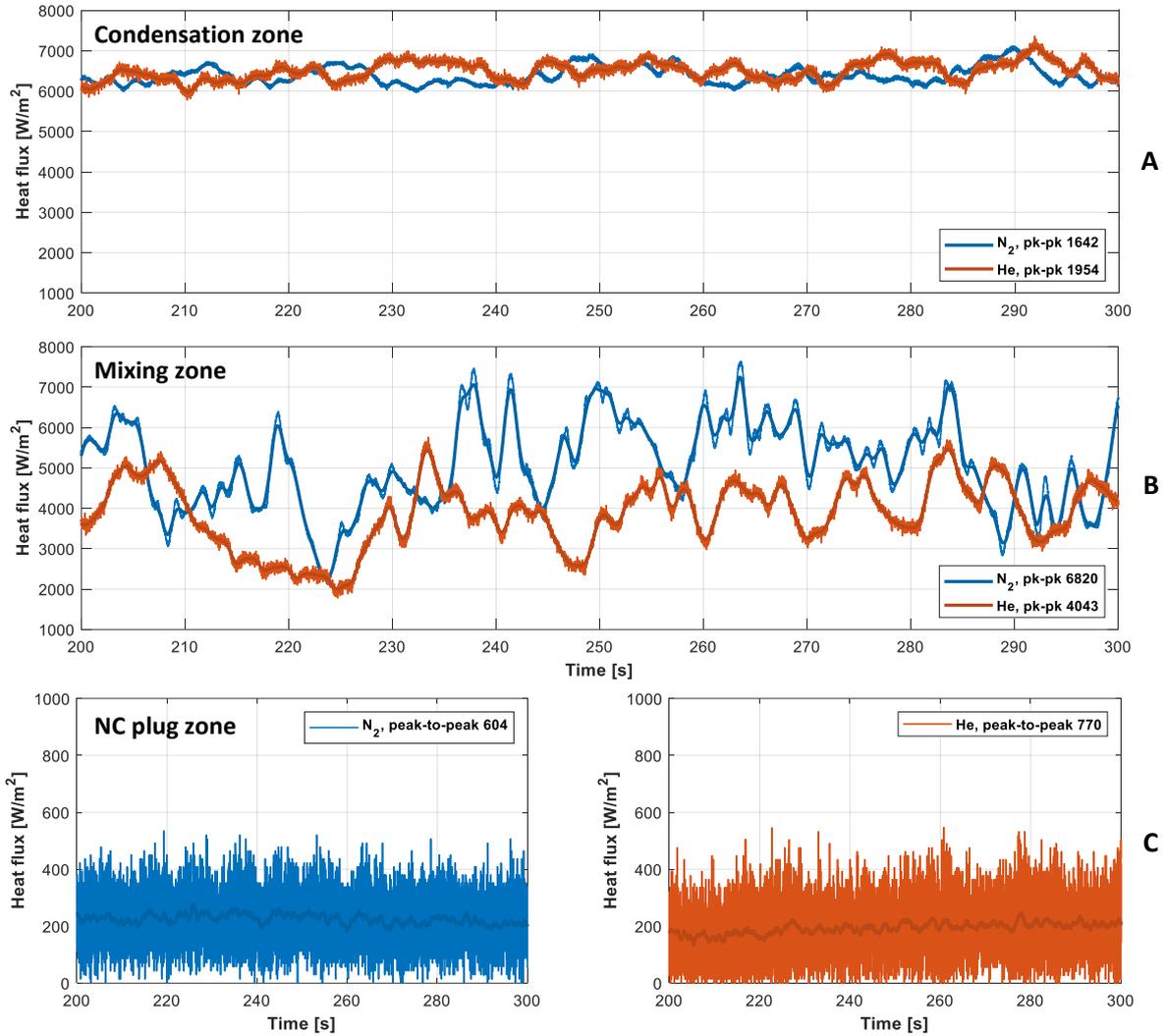

Figure 8 Heat flux signal measured in the condensation zone (A), mixing region (B) and gas plug (C) for experiments with N2 and He. Figure reproduced from [21].

Furthermore, to validate the signal observed with heat flux sensors, a comparison was made against values calculated based on the wall ΔT measured with calibrated K-type thermocouples, eq.(9):

$$\dot{Q} = k\nabla T \qquad (9)$$

where $k$ is the thermal conductivity of wall material and $\nabla T$ is the temperature gradient, which can be defined here as:

$$\nabla T = \frac{\Delta T}{\Delta x} \qquad (10)$$

Figure 9A presents the signal obtained in the condensation zone, while Figure 9B in the mixing zone. In both methods, the increase in signal oscillations is clearly discernible. The slight discrepancy between two methods can be attributed to calibration errors.

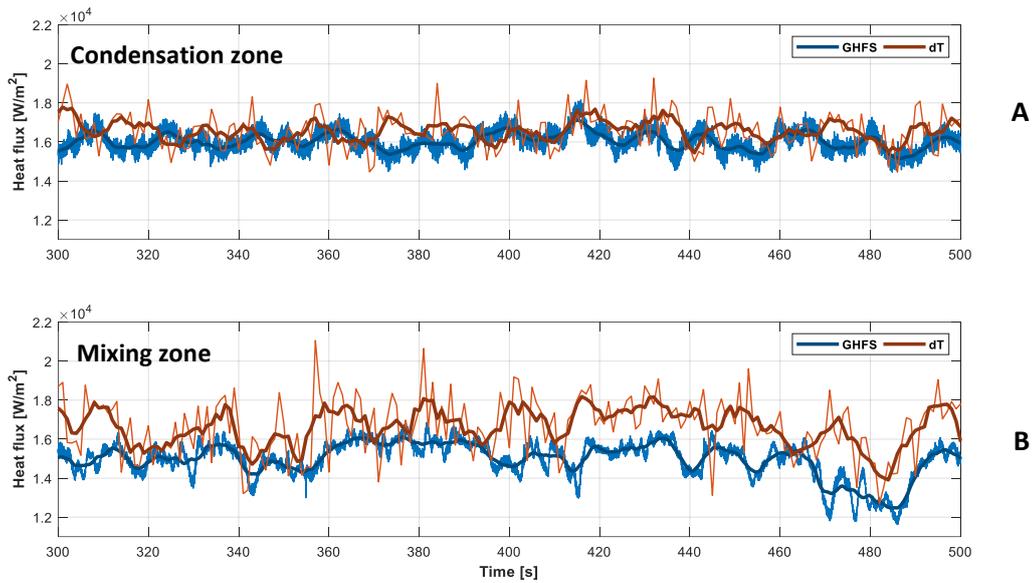

**Figure 9 Comparison between directly measured heat flux signal and values recalculated from ΔT measurement**

Figure 10 presents heat flux measured with a sensor located at 620 mm elevation for experiments with variable inventory of NC gases. X axis is displayed in normalized elevation coordinates, used to superimpose the data from multiple experiments. Position -1 corresponds to the interface between zones A&B, 0 is the zone B center and +1 is the interface between zones B&C. The experiments had varied amount of added NC gas which impacted the size of the NC gas plug at the top of the condensation tube and effectively controlled the vertical position of the mixing zone. Moreover, NC mixture composition was changed between pure $N_2$, He or 1:1 mixture of both – parameter which impacted the mixing zone size. Applied normalization removed the effect of mixing zone position and size and the heat flux profile could have been reconstructed with a use of a single sensor. Remarkably, after the data collapse, the shape of the profile did not vary for different NC gas mixture compositions, even though in the absolute coordinates, the width of the mixing zone varied significantly.

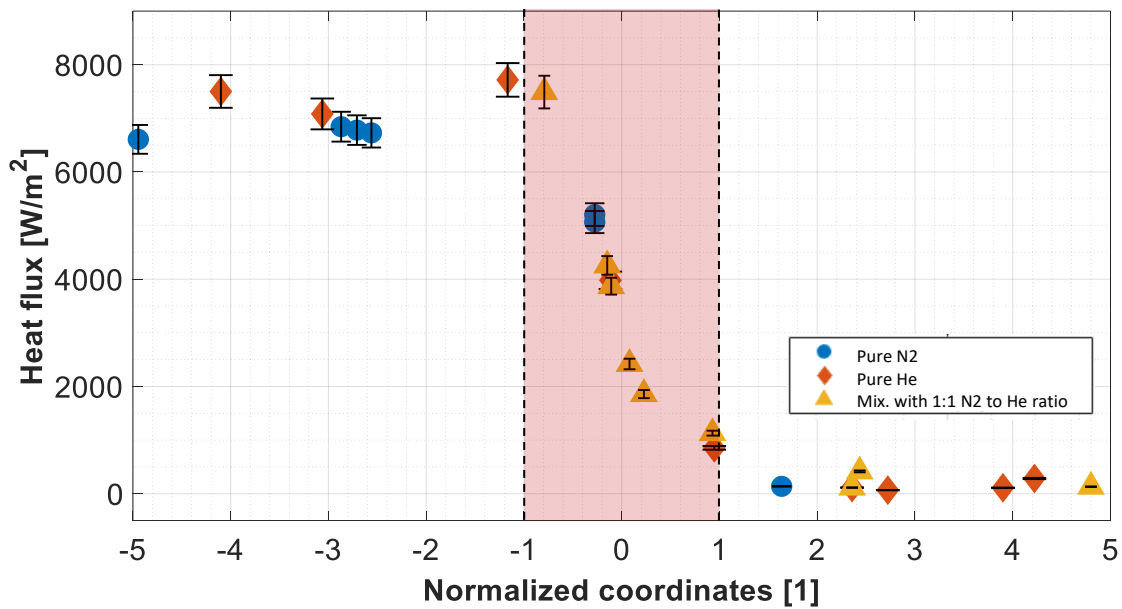

**Figure 10 Heat flux observed in normalized coordinates, with respect to the position of the sensor (620 mm) and the mixing zone center. Figure reproduced from [21].**

## 6.2 Continuous injection tests

To achieve greater spatial resolution of the mixing layer, second type of experiments was conducted in which at the beginning, steady-state in the condensation tube was reached without addition of any NC gases. After confirming the pressure stability, a valve was opened, connecting the test volume to the separate tank with pressurized NC gas. The overpressure and a check-valve in the line ensured flow from the tank into the primary loop. For each NC gas mixture, two overpressures were tested – 0.2 and 0.5 bar. The mass flow of the gas was verified with temperature and pressure drop observation in the NC tank. The gas was introduced to the steam generation tank, from which it was carried upwards to the condensation tube and collected at the top creating a plug. With continuous injection of additional gas, the plug expanded downwards, and the mixing zone moved with it, passing along fixed GHFS's. The pressure difference between the condensation volume and NC gas tank was adjusted so that the zones downward velocity was reasonably low (1-2 mm/s), allowing good spatial as well as temporal resolution of the heat flux vertical profiles. For pure $N_2$, He and mixture of the two the same flow rates were maintained. With the higher overpressure, the NC plug expansion was considerably faster, but remaining characteristics of the zone (temperature/heat flux oscillations) weren't affected. Data presented in this section was collected at 4 bar pressure, 20 °C wall ΔT and 2 m³/h coolant water flow.

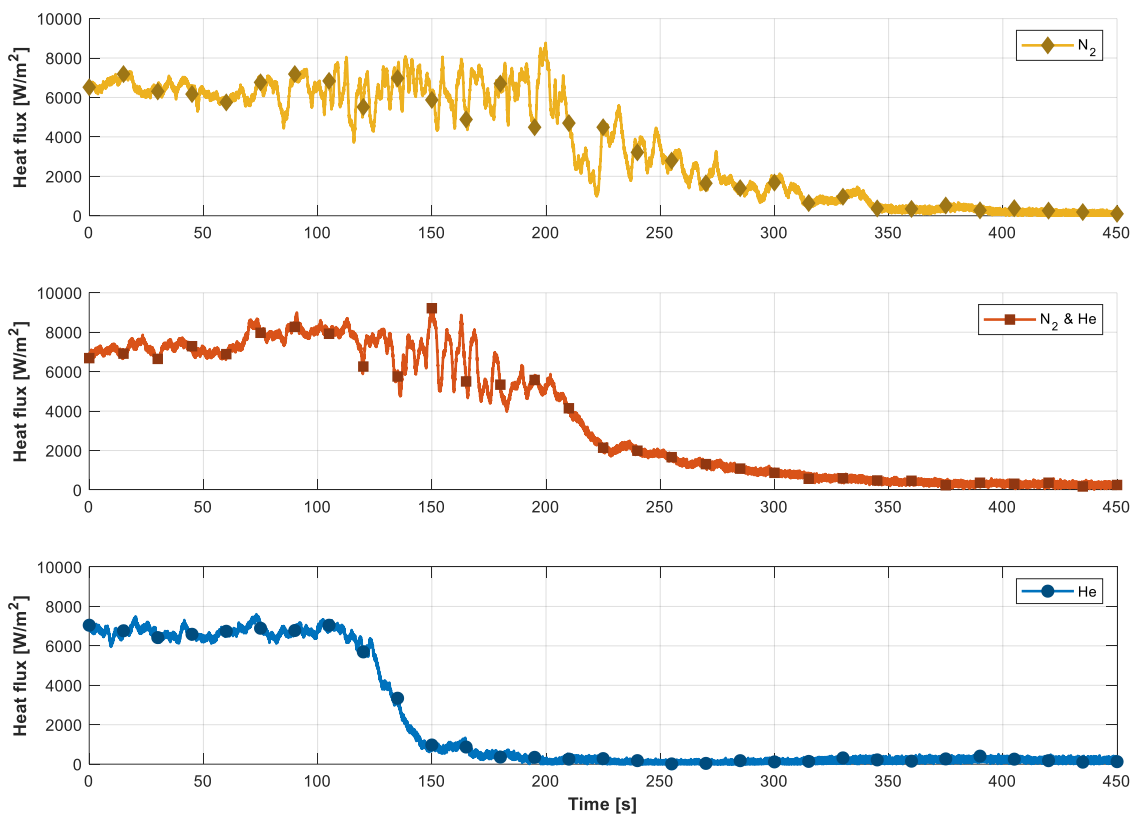

**Figure 11 Mixing zone passage recorded by GHFS at 620 mm for three injected NC mixtures.**

The difference of the vertical heat flux distribution and its oscillation depending on the injected NC gas is clearly distinguishable, Figure 11. Nitrogen, being heavier than steam, promotes mixing in the intermediate zone, resulting in the greatest size as well as variation. Light helium on the other hand produces sharp interface with very little turbulence observable. The NC gases mixture, which was composed of 1:1 by moles combination of $N_2$ and He, exhibits intermediate behavior.

## 7 CONCLUSIONS AND OUTLOOK

Experiments conducted in PRECISE facility confirmed that GHFS's are a useful tool in characterizing condensation heat flux in geometry of SG tubes operating in reflux condenser mode. Time-averaged as well as transient behavior was well resolved, allowing investigation up to 10 Hz frequency when sensors were mounted behind a protective steel layer.

Due to GHFS small dimensions and thermal properties, only small disturbance was introduced into the heat conduction path from steam to coolant water. As a result, the previously observed gas stratification in the condensation tube could be accurately resolved and observed in heat flux dimension. Significant differences in the observed signal allowed for reliable identification of the type of the zone at each GHFS elevation. Moreover, analysis of heat flux transients revealed differences observable for different species of NC gases mixing with steam.

The drawback of the currently installed GHFSs in PRECISE facility is their low V/W sensitivity, which necessitated inclusion of amplification in the measurement chain. Future implementations of this type of sensors should consider other material compositions, to amend this issue, should the test conditions be less harsh.

Stainless-steel / nickel composite GHFS's proved robust to withstand conditions expected during RHR system loss in mid-loop operation of a typical PWR reactor. However, due to the employed embedding technique, their potential to investigate higher-frequency phenomena was not fully utilized in PRECISE facility experiments. Future development in this direction could potentially remove the use of the fragile epoxy and thus the need for the assembly protection. Depending on the investigated phenomena, the fast response time of the fully exposed sensor can provide otherwise unattainable insights.

## 8 ACKNOWLEDGEMENTS
The authors would like to express their sincerest gratitude for to Dr. Terttaliisa Lind whose expertise and guidance greatly assisted the research and helped to improve this manuscript significantly. We are also grateful to Mr. Marton Szogradi for his assistance with conducting the experiments. We have to express our appreciation to the laboratory staff for helping to design and construct the experimental facility.

1. D'Auria, F. & Frogheri, M. Use of a natural circulation map for assessing PWR performance. *Nucl. Eng. Des.* **215**, 111–126 (2002).
2. Yong Jeong, H., Nyun Kim, B. & Lee, K. Thermal-hydraulic phenomena during reflux condensation cooling in steam generator tubes. *Ann. Nucl. Energy* **25**, 1419–1428 (1998).
3. Kawanishi, K., Tsuge, A., Fujiwara, M., Kohriyama, T. & Nagumo, H. Experimental study on heat removal during cold leg small break LOCAs in PWRs. *J. Nucl. Sci. Technol.* **28**, 555–569 (1991).
4. Chatterjee, B. *et al.* Analyses for VVER-1000 / 320 reactor for spectrum of break sizes along with SBO. *Ann. Nucl. Energy* **37**, 359–370 (2010).
5. Kauppinen, O., Kouhia, V., Riikonen, V., Hyvärinen, J. & Sjövall, H. Computer analyses on loop seal clearing experiment at PWR PACTEL. *Ann. Nucl. Energy* **85**, 47–57 (2015).
6. Wang, W. W., Su, G. H., Qiu, S. Z. & Tian, W. X. Thermal hydraulic phenomena related to small break LOCAs in AP1000. *Prog. Nucl. Energy* **53**, 407–419 (2011).
7. Kim, Y. & Kang, K. Overview of an investigation into SBLOCA tests of ATLAS facility. *Ann. Nucl. Energy* **102**, 386–401 (2017).
8. Freixa, J., Reventós, F., Pretel, C., Batet, L. & Sol, I. SBLOCA with boron dilution in pressurized water reactors . Impact on operation and safety. *Nucl. Eng. Des.* **239**, 749–760 (2009).
9. Birchley, J., Haste, T. J. & Richner, M. Accident management following loss of residual heat removal during mid-loop operation in a Westinghouse two-loop PWR. *Nucl. Eng. Des.* **238**, 2173–2181 (2008).
10. Dumont, D., Lavialle, G., Noel, B. & Deruaz, R. Loss of residual heat removal during mid-loop operatio : BETHSY experiments. *Nucl. Eng. Des.* **149**, 365–374 (1994).
11. Fletcher, C., McHugh, P., Naff, S. & Johnsen, G. *Thermal-hydraulic processes involved in loss of residual heat removal during reduced inventory operation. Revision 1*. http://www.osti.gov/energycitations/product.biblio.jsp?osti_id=10103438 (1991).
12. Kral, P. & Rez, U. J. V. Sources and Effect of Non - Condensable Gases in Reactor Coolant System of Lwr. *Nureth-16* 5194–5208 (2015).
13. Mityakov, A. V. *et al.* Gradient heat flux sensors for high temperature environments. *Sensors Actuators A Phys.* **176**, 1–9 (2012).
14. Sapozhnikov, S. Z., Mityakov, V. Y. & Mityakov, a. V. Gradient heat-flux sensors: Possibilities and prospects of use. *Therm. Eng.* **53**, 270–278 (2006).
15. Kyarad, A. & Lengfellner, H. Al-Si multilayers: A synthetic material with large thermoelectric anisotropy. *Appl. Phys. Lett.* **85**, 5613–5615 (2004).
16. Fischer, K., Stoiber, C., Kyarad, A. & Lengfellner, H. Anisotropic thermopower in tilted metallic multilayer structures. *Appl. Phys. A Mater. Sci. Process.* **78**, 323–326 (2004).
17. Zahner, T., Förg, R. & Lengfellner, H. Transverse thermoelectric response of a tilted metallic multilayer structure. *Appl. Phys. Lett.* **73**, 1364–1366 (1998).
18. Cardarelli, F. *Material Handbook - A Concise Desktop Reference*. (Springer Handbook, 2008). doi:10.1007/978-1-84628-669-8.
19. Raphael-Mabel, S. Design and Calibration of a Novel High Temperature Heat Flux Sensor. (2005).
20. Janasz, F., Prasser, H.-M. H.-M., Suckow, D. & Szogradi, M. Non-Condensable Gas Plugging and Mixing Behavior in PWR Steam Generator Tubes During Reflux Condensation. in *26th International Conference on Nuclear Engineering* vol. Volume 6B: V06BT08A051 (ASME, 2018).
21. Janasz, F. Effect of non-condensable gases on reflux condensation in nuclear steam generator tubes. (ETH Zurich, 2019). doi:10.3929/ETHZ-B-000387773.